\documentclass[conference]{IEEEtran}
\IEEEoverridecommandlockouts

\usepackage{cite}
\usepackage{amsmath,amssymb,amsfonts}
\usepackage{algorithmic}
\usepackage{graphicx}
\usepackage{textcomp}
\usepackage{xcolor}

\usepackage[T1]{fontenc}
\usepackage{booktabs}
\usepackage[misc]{ifsym}

\usepackage{dsfont}
\usepackage{amsmath,amssymb,amsfonts}
\usepackage{hyperref}

\def\BibTeX{{\rm B\kern-.05em{\sc i\kern-.025em b}\kern-.08em
    T\kern-.1667em\lower.7ex\hbox{E}\kern-.125emX}}
\begin{document}

\title{Probabilistic Forecasting Cryptocurrencies Volatility: From Point to Quantile Forecasts  }

\author{
\IEEEauthorblockN{Grzegorz Dudek}
\IEEEauthorblockA{\textit{Faculty of Electrical Engineering} \\
\textit{Częstochowa University of Technology}\\
Częstochowa, Poland \\
grzegorz.dudek@pcz.pl}
\and
\IEEEauthorblockN{Witold Orzeszko}
\IEEEauthorblockA{\textit{Faculty of Economic Sciences} \\
\IEEEauthorblockA{\textit{and Management}} 
\textit{Nicolaus Copernicus University in Toruń}\\
Toruń, Poland \\
witold.orzeszko@umk.pl}
\and
\IEEEauthorblockN{Piotr Fiszeder}
\IEEEauthorblockA{\textit{Faculty of Economic Sciences} \\
\textit{and Management} \\
\textit{Nicolaus Copernicus University in Toruń}\\
Toruń, Poland \\
\textit{Faculty of Finance and Accounting} \\
\textit{Prague University of Economics}\\
\textit{and Business}\\
Prague, Czech Republic \\
piotr.fiszeder@umk.pl}
}

\maketitle

\begin{abstract}
Cryptocurrency markets are characterized by extreme volatility, making accurate forecasts essential for effective risk management and informed trading strategies. Traditional deterministic (point) forecasting methods are inadequate for capturing the full spectrum of potential volatility outcomes, underscoring the importance of probabilistic approaches. To address this limitation, this paper introduces probabilistic forecasting methods that leverage point forecasts from a wide range of base models, including statistical (HAR, GARCH, ARFIMA) and machine learning (e.g. LASSO, SVR, MLP, Random Forest, LSTM) algorithms, to estimate conditional quantiles of cryptocurrency realized variance. 
To the best of our knowledge, this is the first study in the literature to propose and systematically evaluate probabilistic forecasts of variance in cryptocurrency markets based on predictions derived from multiple base models. Our empirical results for Bitcoin demonstrate that the Quantile Estimation through Residual Simulation (QRS) method, particularly when applied to linear base models operating on log-transformed realized volatility data, consistently outperforms more sophisticated alternatives. Additionally, we highlight the robustness of the probabilistic stacking framework, providing comprehensive insights into uncertainty and risk inherent in cryptocurrency volatility forecasting. This research fills a significant gap in the literature, contributing practical probabilistic forecasting methodologies tailored specifically to cryptocurrency markets.
\end{abstract}


\section{Introduction}

Probabilistic forecasting of cryptocurrency volatility is essential due to the considerable uncertainty and frequent occurrence of extreme price movements in cryptocurrency markets. Unlike traditional point forecasts, probabilistic methods estimate the entire conditional distribution (or its fine-grained approximation using densely spaced quantiles) of future volatility, thereby capturing the full range of potential outcomes and significantly improving risk assessment and decision-making in these highly unpredictable markets.

Despite these clear benefits, probabilistic forecasting methods remain relatively scarce in the cryptocurrency volatility literature. To address this gap, our study proposes novel approaches for generating probabilistic forecasts of cryptocurrency realized variance, introducing quantile-based methods that leverage deterministic forecasts from multiple base models to produce comprehensive, uncertainty-aware predictions.

\subsection{Related Work}

Forecasting cryptocurrency volatility is particularly challenging due to extreme price fluctuations, frequent outliers, market microstructure effects, diverse investor time scales, speculative bubble behaviors, and strong sensitivity to market dynamics \cite{Bou19}. Traditional methods such as GARCH (Generalized Autoregressive Conditional Heteroscedasticity), HAR (Heterogeneous Autoregressive), and ARFIMA (Autoregressive Fractionally Integrated Moving Average) are widely used but have inherent limitations, including their linear assumptions and limited adaptability to rapidly evolving cryptocurrency markets \cite{Ber22}. To address these shortcomings, machine learning (ML) methods have emerged as promising alternatives \cite{Sez20}, \cite{Dud24a}, offering enhanced flexibility, nonlinearity, and adaptability. Their predictive power can be further improved through the integration of time series decomposition, advanced optimization techniques, ensemble learning \cite{Nad25}, and hybridization with classical statistical approaches \cite{Kri18}.

Despite these advancements, most cryptocurrency volatility forecasting methods remain deterministic, focusing on point forecasts rather than full probability distributions. Probabilistic forecasting approaches, which estimate the entire distribution of possible outcomes or its approximation using quantile estimates, are significantly less common. One example is \cite{Gol24}, where the authors proposed a model incorporating probabilistic gated recurrent units to generate probability distributions for predicted values. However, despite the model’s probabilistic nature, the study focused solely on point forecasts without evaluating the quality of the probabilistic predictions. Another approach, introduced in \cite{Hon24}, employs a variational autoencoder learning framework for multivariate distributional forecasting. This method directly estimates the cumulative distribution function (CDF) of future time series conditional distributions, enabling probabilistic forecasting by generating synthetic time series data for future time points.



A unique approach to probabilistic forecasting based on point forecasts was proposed in \cite{Dud24}. The study introduced probabilistic stacking, an extension of traditional stacking, where multiple base forecasts are combined using a meta-model designed to produce quantile-based probabilistic predictions. Stacking is well known in its deterministic form and has been previously applied to cryptocurrency volatility forecasting \cite{Ara21}. This methodology enhances predictive accuracy by combining multiple models, leveraging their complementary strengths, and reducing overfitting. It also increases model diversity, captures nonlinear relationships through a meta-learner, and balances biases, making it a robust and adaptable approach for forecasting. Probabilistic stacking extends these advantages further by generating full probabilistic forecasts in the form of quantiles, providing a richer representation of uncertainty. This study explores the application of probabilistic stacking to cryptocurrency volatility forecasting, demonstrating its effectiveness in capturing market uncertainty.

\subsection{Motivation and Contributions}

Motivated by the substantial volatility and unpredictable price movements inherent in cryptocurrency markets, and recognizing the limitations of deterministic forecasts, this study addresses the scarcity of probabilistic forecasting methods in cryptocurrency volatility research by proposing and evaluating novel quantile-based forecasting approaches.
Our key contributions include:

\begin{itemize}
    \item \textbf{Conversion of point forecasts into probabilistic forecasts:} We propose approaches that transform deterministic point forecasts of cryptocurrency volatility into probabilistic (quantile-based) forecasts, capturing the inherent uncertainty in cryptocurrency markets.

    \item \textbf{Probabilistic stacking:} We extend the classical deterministic stacking approach to its probabilistic variant for probabilistic forecasting of cryptocurrency variance. We employ Quantile Linear Regression and Quantile Regression Forests as meta-models to produce quantile-based forecasts.
    \item \textbf{Integration of diverse base models:} We leverage twelve base models, including classical statistical methods and modern ML algorithms, to generate point forecasts of cryptocurrency volatility, subsequently used as inputs for our probabilistic models.
    \item \textbf{Comprehensive evaluation framework:} We systematically evaluate and compare the probabilistic forecasting methods using robust metrics: the Continuous Ranked Probability Score, Relative Frequency, and Winkler Score.
    \item \textbf{Analytical insights:} 
    Our analysis offers detailed insights into the strengths and limitations of each probabilistic forecasting method in the context of cryptocurrency volatility.
\end{itemize}

We test the proposed methodology using Bitcoin data, but we believe it can also be effectively applied to other cryptocurrencies, as their volatility time series share similar characteristics (see Fig. 2 in \cite{Dud24a}).

The rest of the paper is organized as follows. Section II describes the dataset and formulates the forecasting problem. Section III outlines the proposed probabilistic forecasting methods. Section IV presents the experimental setup and results. Section V provides a discussion of the findings. Section VI concludes the paper and suggests directions for future work.


\section{Data and Problem Statement}

In the experimental part of this study, we test the proposed approaches using Bitcoin (BTC/USD) data. The dataset was obtained from the crypto exchange Kraken and covers the period from 2017 to 2021. 

For each day $t$, we estimate the daily realized variance (RV), denoted as $RV_d$:

\begin{equation}
RV_{d,t} = \sum_{k=1}^{K} r_{k,t}^2, \qquad r_{k,t} = \ln P_{k,t} - \ln P_{k-1,t}
\end{equation}
where $K$ is the number of intraday return observations within a day (288 for 5-minute returns in our case), $r_{k,t}$ represents the $k$-th intraday return on day $t$, and $P_{k,t}$ is the price of Bitcoin at the $k$-th observation within day $t$. 

\begin{figure}[t]
\centering
\includegraphics[width=0.24\textwidth]{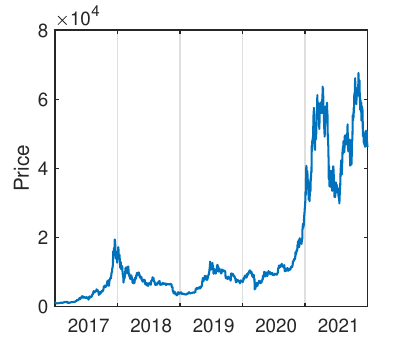}
\includegraphics[width=0.24\textwidth]{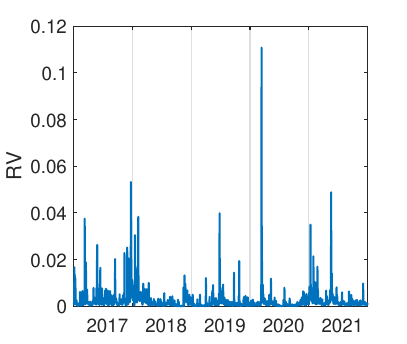}
\includegraphics[width=0.24\textwidth]{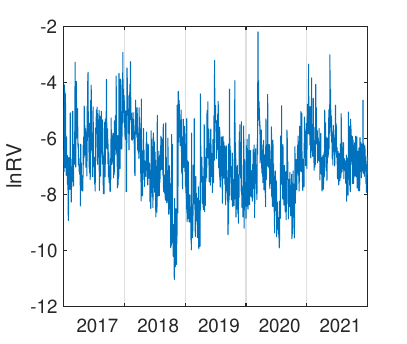}
\caption{BTC/USD data: price, $RV_d$, and $\ln RV_d$.}  
\label{figDT}
\end{figure}

The plot of $RV_d$ exhibits multiple spikes, as shown in Fig.~\ref{figDT}. These outliers pose a challenge for forecasting models. Thus, forecasting models often operate on log-transformed data, $\ln RV_d$, which is presented in the right panel of Fig.~\ref{figDT}.  

Probabilistic forecasting of cryptocurrencies RV aims to estimate the entire distribution of future volatility rather than providing a single-point prediction. In the framework of quantile forecasting considered in this study, the objective is to predict conditional quantiles of RV based on point forecasts from base models.  

Given point forecasts $\hat{\textbf{y}}_t = [\hat{y}_{1,t}, ..., \hat{y}_{n,t}]$ from $n$ base models, the quantile forecasting method estimates the conditional quantile of level $q$ using a regression function (meta-model): $\hat{y}_{q,t} = f(\hat{\textbf{y}}_t; \boldsymbol{\theta}_q)$
where $\boldsymbol{\theta}_q$ denotes the model parameters. Some models, such as Quantile Regression Forests, enable the simultaneous prediction of multiple quantiles: $\hat{\mathbf{y}}_{\Pi,t} = f(\hat{\textbf{y}}_t; \boldsymbol{\theta})$, where $\hat{\mathbf{y}}_{\Pi,t}=[\hat{y}_{q,t}]_{q \in \Pi}$.  

The class of regression functions $f$ encompasses a wide range of mappings, including both linear and nonlinear models. Their parameters may remain static or evolve over time. To enhance the performance of the meta-model, we adopt an approach where the parameters are learned individually for each forecasting task (forecasted day $\tau$) using training set $\Phi = \{(\hat{\textbf{y}}_t, y_t)\}_{t \in \Psi}$, where $y_t$ represents the target value, and $\Psi=\{1, ..., \tau-1\}$.  

We consider two versions of meta-models. The first operates on log-transformed data ($\ln RV_d$), while the second uses raw data ($RV_d$). In the experimental section, we denote the first version with the suffix '-l'.  

To improve the performance of the meta-model, we incorporate supplementary inputs. These inputs are the same variables used to train the base models \cite{Dud24a}: $\mathbf{x}_t = [ \ln RV_{d,t-1}, \ln RV_{w,t-1}, \ln RV_{m,t-1}]$ (or $\mathbf{x}_t = [RV_{d,t-1}, RV_{w,t-1}, RV_{m,t-1}]$ in the raw data variant), where $RV_w=\frac{1}{7}\sum_{i=1}^{7}RV_{d,t-i}$ denotes the weekly RV, and $RV_w=\frac{1}{30}\sum_{i=1}^{30}RV_{d,t-i}$ 
represents the monthly RV.  
The extended models take the following form: $\hat{y}_{q,t} = f(\hat{\textbf{y}}_t, \mathbf{x}_t; \boldsymbol{\theta}_q)$ ($\hat{\mathbf{y}}_{\Pi,t} = f(\hat{\textbf{y}}_t, \mathbf{x}_t; \boldsymbol{\theta})$). In the experimental section, these models are denoted with the suffix '-e'.  

\section{Probabilistic Forecasting Approaches}

In our research, we introduce three approaches for probabilistic forecasting of cryptocurrency volatility. Each approach employs a distinct mechanism to model the distribution of next-day forecasts in the form of 99 quantiles ($q \in \Pi = {0.01, 0.02, ..., 0.99}$), based on point forecasts from the base models.

\subsection{Quantile Estimation through Residual Simulation (QRS)}

The residual simulation approach for quantile estimation is a method used in probabilistic forecasting to construct predictive quantiles by leveraging the residuals of a base predictive model. This approach assumes that the distribution of forecast errors (residuals) from historical data can be used to estimate the uncertainty of future predictions.  

Below is a step-by-step description of this approach. Suppose our objective is to generate a quantile forecast for day $\tau$, given the point forecast for that day, denoted as $\hat{y}_{\tau}$. The distribution of forecast errors is estimated based on past point forecasts from the preceding period, defined as $\Psi = \{1, ..., \tau-1\}$.  

\begin{enumerate}
    \item \textbf{Base model forecasting.} A deterministic (point) forecast model is trained on historical data to generate predictions, following the methodology described in \cite{Dud24a}. The model provides point estimates $\hat{y}_t$ for the period $\Psi$ as well as for $\tau$.  

    \item \textbf{Residual calculation.} The residuals are computed as:  
    \begin{equation}
        e_t = y_t - \hat{y}_t,\quad t \in \Psi
    \end{equation}  
    These residuals represent the model’s forecasting errors and serve as an empirical measure of prediction uncertainty.  

    \item \textbf{Simulated forecast distribution.} The simulated forecasts are obtained by adding the residuals to the base model’s prediction for day $\tau$:  
    \begin{equation}
        \tilde{y}_\tau = \hat{y}_\tau + e_t,\quad t \in \Psi
    \end{equation}  
    This results in an empirical distribution of potential future values:  
    \begin{equation}
        \Lambda = \{\hat{y}_\tau + e_1, \hat{y}_\tau + e_2, ..., \hat{y}_\tau + e_{\tau-1}\}
    \end{equation}  

    \item \textbf{Distribution function fitting.} A distribution function is fitted to $\Lambda$. We use a nonparametric kernel density estimation approach, which is more flexible than parametric alternatives. Specifically, we apply a Gaussian kernel with bandwidth optimized for normal densities.  

    \item \textbf{Quantile estimation.} The desired quantiles are then extracted from the fitted distribution using the inverse CDF.  
\end{enumerate}

This approach offers several advantages. It is data-driven and model-agnostic, meaning it can be applied to any forecasting model without requiring modifications to its structure. Additionally, it is computationally efficient, as it is relatively simple to implement and requires less computational power compared to more complex probabilistic models.  

However, the method also has some limitations. It assumes that past residuals accurately represent future uncertainty, which may not hold under volatile conditions. Furthermore, if the base model exhibits systematic bias, the estimated quantiles may inherit this bias. Lastly, it does not account for time-varying changes in the error distribution, limiting its adaptability. 
In our implementation, we attempt to mitigate this limitation by extending the set $\Lambda$ up to the forecasted day $\tau$.

\subsection{Quantile Linear Regression (QLR)}

QLR is a statistical method used for estimating conditional quantiles of a response variable, given a set of predictor variables. Unlike ordinary least squares regression (OLS), which estimates the mean of the response variable conditional on the predictors, QLR estimates a specific quantile of the conditional distribution. This makes it especially useful for cases where the relationship between the response and the predictors may differ across different points of the distribution, such as when modeling asymmetric or heteroscedastic data.

As noted by Koenker in \cite{Koe01}, a linear model given by

\begin{equation}
 f(\hat{\textbf{y}}) = \sum_{i=1}^n{a_i\hat{y}_i} + a_0
 \label{eq1}
\end{equation}
where $n$ is the number of base models and $a_0, ..., a_n$ are coefficients,
can effectively estimate quantiles when optimized using the pinball loss function:

\begin{equation}
L_q(y, \hat{y}_{q}) =
\begin{cases}
(y-\hat{y}_{q})q & \text{if } y \geq \hat{y}_{q}\\
(y-\hat{y}_{q})(q-1)  &\text{if } y < \hat{y}_{q} 
\end{cases}
\label{eqrho}
\end{equation}
where $y$ represents the true value, and $\hat{y}_{q}$ is its predicted $q$-quantile.

Unlike OLS, which minimizes the sum of squared errors, QLR minimizes the sum of absolute errors with weights depending on the quantile. This allows the model to focus on different parts of the distribution, providing a more flexible and robust approach to regression when dealing with non-normal data or outliers. The optimization problem remains linear and can be efficiently solved using the interior point (Frisch-Newton) algorithm.

Compared to OLS, QLR is less sensitive to extreme values that could otherwise distort regression estimates. Its ability to model multiple quantiles provides a richer understanding of the data distribution, making it particularly valuable in risk analysis applications in economics and finance.

However, QLR has some limitations. It assumes a linear relationship between predictors and quantiles, which may not hold for highly nonlinear data, potentially leading to inaccurate estimates. Moreover, while it is less computationally intensive than more complex probabilistic models, it can still be demanding, particularly when estimating multiple quantiles, as a separate model must be trained for each probability level $q$. This computational burden becomes even more pronounced when dealing with large datasets.

\subsection{Quantile Regression Forest (QRF)}

QRF is an extension of the Random Forest (RF) algorithm designed for probabilistic regression. Unlike standard RF, which provides only a mean prediction, QRF estimates the entire conditional distribution of the response variable, allowing for the computation of predictive quantiles. This makes QRF particularly useful for uncertainty estimation and risk-sensitive applications.

QRF follows the same structure as RF, consisting of an ensemble of decision trees trained on bootstrap samples of the dataset. Each tree is grown using a recursive partitioning approach, splitting the data based on predictor variables to minimize impurity. However, unlike RF, which averages the predictions of terminal nodes to obtain a point estimate, QRF retains the full distribution of observed response values within each leaf node.
This distribution is approximated in QRF as follows \cite{Mei06}:

\begin{equation}
 \hat{F}(y|X = \hat{\textbf{y}})=
\sum_{t \in \Psi}w_{t}(\hat{\textbf{y}})\mathds{1}{\{y_{t} \leq y\}} 
\label{eqdt2}
\end{equation}
\begin{equation}
 w_{t}(\hat{\textbf{y}})=\frac{1}{p}\sum_{j=1}^p
 \frac{\mathds{1}{\{\hat{\textbf{y}}_{t} \in \ell_j(\hat{\textbf{y}})\}}}
 {\sum_{\kappa \in \Psi}\mathds{1}{\{\hat{\textbf{y}}_{\kappa} \in \ell_{j}(\hat{\textbf{y}})\}}} 
\label{eqdt2a}
\end{equation}
where the weights $w_t$ determines the contribution of each training sample to the empirical CDF, $p$ is the number of trees in the forest, 
$\ell_{j}$ denotes the leaf that is obtained when dropping $\hat{\textbf{y}}$ down the $j$-th tree, and $\mathds{1}$ is the indicator function.

This empirical distribution is then used to estimate predictive quantiles.

QRF has many advantages. It is nonparametric and can model complex, nonlinear relationships without making explicit distributional assumptions. Additionally, since it captures the full conditional distribution of the response variable, it provides a flexible and robust method for quantile estimation, making it well-suited for high-dimensional or heteroscedastic data.

However, QRF has some drawbacks. It can be computationally expensive, particularly for large datasets, as it requires storing and analyzing the full distribution of responses within each leaf. Additionally, QRF may be less effective when dealing with small sample sizes, as its quantile estimates rely on the empirical distribution of the available data points, which may not be well-represented. Moreover, QRF requires tuning of several hyperparameters for optimal performance, including the number of trees, the number of randomly selected predictors at each decision split, and the minimum number of samples per leaf. The last of these, which determines tree depth, is particularly important for managing the bias-variance tradeoff of the estimator.

\section{Experimental Study}

In this section, we compare the performance of QRS, QLR, and QRF in probabilistic forecasting of cryptocurrency volatility using Bitcoin as a case study.

\subsection{Base Models}

The base models employed in this study are adopted from \cite{Dud24a}, where they were initially proposed for point forecasting of cryptocurrency RV. These models comprise a diverse set of classical statistical methods as well as modern machine learning algorithms: \\
\textbf{HAR} -- Heterogeneous AutoRegressive model,\\
\textbf{HAR-R} -- Heterogeneous AutoRegressive model with Robust estimation,\\
\textbf{ARFIMA} -- AutoRegressive Fractionally Integrated Moving Average model,\\
\textbf{GARCH} -- Generalized AutoRegressive Conditional Heteroscedasticity model,\\
\textbf{LASSO} -- Least Absolute Shrinkage and Selection Operator model,\\
\textbf{RR} -- Ridge Regression,\\
\textbf{SVR-G} -- Support Vector Regression with a Gaussian kernel,\\
\textbf{SVR-L} -- Support Vector Regression with a Linear kernel,\\
\textbf{MLP} -- Multi-Layer Perceptron neural network,\\
\textbf{FNM} -- Fuzzy Neighborhood Model,\\
\textbf{RF} -- Random Forest,\\
\textbf{LSTM} -- Long Short-Term Memory neural network.

\subsection{Training and Optimization Setup}

The base models generated one-step-ahead point forecasts of RV for each day over a three-year period (2019–2021). All base models were appropriately optimized, with the optimization procedure and training process described in detail in \cite{Dud24a}.

The point forecasts from the base models in the final year, 2021, serve as the test set for the probabilistic models. These probabilistic models are trained independently for each sample in the test set, meaning they are trained 365 times. The training sets contain all prior observations (i.e., the base model forecasts from the beginning of 2019 up to the day preceding the forecast). After training, the probabilistic models predict 99 quantiles ($q \in \Pi = {0.01, 0.02, ..., 0.99}$) for each day in the test period.

QRS calculates quantiles for each base model based on its past forecasts, repeating this process for each day in the test period. Additionally, two ensemble solutions are introduced: the mean of the base models (Ens-Mean) and the median of the base models (Ens-Med). QRS also generates probabilistic forecasts for these ensembles in the same manner as for the individual base models.

Two QRS variants are considered: one operates on the logarithm of RV, while the other operates on raw RV values. The former variant is denoted with the suffix '-l' (e.g., HAR-l, LSTM-l).

For QRS, quantiles were calculated using the inverse CDF. In our Matlab implementation, we used the \texttt{icdf} function for this purpose. However, for QRS operating on raw RV data, this function failed to converge for certain $q$, typically for large $q$ (0.98 and 0.99). These cases accounted for less than 1\% of instances, with the highest occurrence observed for SVR-G ($\sim$ 3.5\%) and LSTM ($\sim$ 1.6\%). To address this issue, we interpolated the missing quantiles based on their neighboring values using a piecewise cubic Hermite interpolating polynomial.

For QLR, we used the Matlab implementation provided by Roger Koenker, solving a linear program via the interior point method (\url{www.econ.uiuc.edu/~roger/research/rq/rq}).

For QRF, we set the number of trees in the forest to 100 and the number of randomly selected predictors for each decision split to $n/3$, following the recommendations of the RF inventors. The only optimized hyperparameter was the minimum number of observations per leaf, which was selected from the set $\{1, 5, 10, ..., 70\}$ using training samples from the 2019–2020 period, with out-of-bag error as the selection criterion.

We analyze four variants of the QLR and QRF models, as detailed in Section 2:
\begin{itemize}
    \item operating on the logarithm of RV, denoted with the suffix '-l',
    \item operating on the logarithm of RV with additional inputs: $\ln RV_d$, $\ln RV_w$, and $\ln RV_m$, denoted with the suffix '-le',
    \item operating on raw RV values, and
    \item operating on raw RV values with additional inputs: $RV_d$, $RV_w$, and $RV_m$, denoted with the suffix '-e'.
\end{itemize}

The proposed meta-models were implemented in Matlab 2023b, and the experiments were conducted on a Microsoft Windows 10 Pro system with an Intel(R) Core(TM) i7-6950X CPU @ 3.0 GHz and 48 GB of RAM.

\subsection{Evaluation Metrics}

The primary evaluation metric used in this study is the Continuous Ranked Probability Score (CRPS), which can be expressed in terms of the pinball loss:

\begin{equation} \label{eqCRPS}
    \text{CRPS}(F, y) \approx \sum_{q \in \Pi} L_{q}(y, \hat{y}_{q})
\end{equation}
where $F$ is the CDF of $y$, $\Pi$ is the set of quantile levels,
$\hat{y}_{q}$ represents the predicted quantile at level $q$, and  
$L_{q}(y, \hat{y}_{q})$ is the pinball loss, as defined in \eqref{eqrho}.

An alternative evaluation of predicted quantiles is based on relative frequency (calibration), defined as follows \cite{Mar22}:

\begin{equation}
\text{ReFr}(q)=\frac{1}{N} \sum_{i=1}^N \mathds{1}{\{y_i \leq \hat{y}_{q,i}\}}
\label{eqrf}
\end{equation}
where $N$ is the number of samples.

The expected value of $\text{ReFr}(q)$ is the nominal probability level 
$q$. In other words, the predicted $q$-quantiles should exceed the realized values in $100q\%$ of cases, ensuring a ReFr equal to 
$q$. To assess the average deviation of $\text{ReFr}(q)$ from the expected 
$q$ across all $q \in \Pi$, we define the Mean Absolute ReFr Error (MARFE) \cite{Dud24}:

\begin{equation}
\text{MARFE}=\frac{1}{|\Pi|} \sum_{q \in \Pi} |\text{ReFr}(q)-q|
\label{eqmrf}
\end{equation}

The Winkler Score (WS) is a proper scoring rule commonly used to evaluate the accuracy of prediction intervals in probabilistic forecasting \cite{Bre21}. It accounts for both the width of the prediction interval and whether the observed value falls within the interval, penalizing predictions that are either too wide or fail to capture the true value. The Winkler Score is defined as:

\begin{equation}
\text{WS} =
\begin{cases}
(\hat{y}_{q_u} - \hat{y}_{q_l}) + \frac{2}{q}(\hat{y}_{q_l} - y) & \text{if } y < \hat{y}_{q_l}\\
(\hat{y}_{q_u} - \hat{y}_{q_l}) & \text{if } \hat{y}_{q_l} \leq y \leq \hat{y}_{q_u}\\
(\hat{y}_{q_u} - \hat{y}_{q_l}) + \frac{2}{q}(y-\hat{y}_{q_u}) & \text{if } y > \hat{y}_{q_u}\\
\end{cases}
\end{equation}
where $\hat{y}_{q_l}$ and $\hat{y}_{q_u}$ represent the predicted lower and upper quantiles defining the $100(1-\alpha)\%$ PI (predictive interval), with $\alpha=q_u-q_l$.

In this study, we evaluate 90\% PIs ($q_l=0.05$ and $q_u=0.95$) using mean WS:

\begin{equation}
\text{MWS} = \frac{1}{N} \sum_{i=1}^{N} \text{WS}(y_i, \hat{y}_{q_l,i}, \hat{y}_{q_u,i})
\label{eqmae}
\end{equation}

To complement the WS metric, we also consider intuitive measures that evaluate the quality of PIs: the percentages of observed values falling within, below, and above the interval. These simple metrics allow for an easy comparison of results against the expected values -- in our case, 90\%, 5\%, and 5\%, respectively.

Using the quantile forecasts, we can also derive a point forecast by assuming the median as the central estimate. For this case, we compute the following point prediction errors:

\begin{equation}
\text{MAE-Q} = \frac{1}{N} \sum_{i=1}^{N} | y_i - \hat{y}_{0.5,i} |
\label{eqmae}
\end{equation}
\begin{equation}
\text{MSE-Q} = \frac{1}{N} \sum_{i=1}^{N} (y_i - \hat{y}_{0.5,i})^2
\label{eqmse}
\end{equation}

By comparing these errors with those obtained from deterministic models, we can assess whether probabilistic forecasting improves point forecast accuracy. 

\subsection{Results}

Table \ref{tab1} summarizes performance metrics for the test data, with the best results highlighted in bold. Due to the stochastic nature of QRF, results shown are averages over 50 training repetitions.

\begin{table*}[htbp]
\begin{center}
\caption{Performance metrics.}\label{tab1}
\begin{tabular}{|l|r|r|r|r|r|r|r|}
\hline
\multicolumn{1}{|c}{Approach/} & \multicolumn{1}{|c}{CRPS} & \multicolumn{1}{|c}{CRPS}   & \multicolumn{1}{|c}{CRPS} & \multicolumn{1}{|c}{MARFE} & \multicolumn{1}{|c}{MWS} & \multicolumn{1}{|c}{MAE-Q} & \multicolumn{1}{|c|}{MSE-Q} \\
Model & \multicolumn{1}{|c}{Mean} & \multicolumn{1}{|c}{Median} & \multicolumn{1}{|c}{IQR}  & \multicolumn{1}{|c}{}      & \multicolumn{1}{|c}{}    & \multicolumn{1}{|c}{}      & \multicolumn{1}{|c|}{}      \\
  \hline
\multicolumn{8}{|l|}{\textbf{QRS-l (logaritmized RV)}}   \\
\hline
HAR-l      & 9.53E-04 & 3.54E-04 & 5.04E-04 & 5.54E-02 & 1.16E-02 & \textbf{1.16E-03} & 1.27E-05 \\
HAR-R-l    & 9.52E-04 & 3.51E-04 & 5.15E-04 & 5.34E-02 & 1.17E-02 & \textbf{1.16E-03} & 1.27E-05 \\
ARFIMA-l   & 9.49E-04 & 3.74E-04 & 4.87E-04 & 4.90E-02 & 1.13E-02 & \textbf{1.16E-03} & 1.25E-05 \\
GARCH-l    & 1.22E-03 & 3.62E-04 & 5.27E-04 & 1.42E-01 & 1.37E-02 & 1.51E-03 & 1.66E-05 \\
LASSO-l    & 1.03E-03 & 3.42E-04 & 4.79E-04 & 1.02E-01 & \textbf{1.11E-02} & 1.30E-03 & 1.40E-05 \\
RR-l       & \textbf{9.45E-04} & 3.56E-04 & 5.28E-04 & 5.23E-02 & 1.13E-02 & \textbf{1.16E-03} & 1.25E-05 \\
SVR-G-l    & 9.53E-04 & 3.77E-04 & 5.41E-04 & 4.68E-02 & 1.14E-02 & 1.17E-03 & 1.22E-05 \\
SVR-L-l    & 9.48E-04 & 3.70E-04 & 5.63E-04 & 4.83E-02 & 1.13E-02 & \textbf{1.16E-03} & 1.22E-05 \\
MLP-l      & 9.62E-04 & 3.78E-04 & 4.93E-04 & 4.91E-02 & 1.14E-02 & 1.20E-03 & 1.20E-05 \\
FNM-l      & 9.71E-04 & 3.61E-04 & 4.95E-04 & 5.77E-02 & 1.15E-02 & 1.20E-03 & 1.30E-05 \\
RF-l       & 9.92E-04 & 3.76E-04 & 5.24E-04 & 5.00E-02 & 1.16E-02 & 1.23E-03 & 1.32E-05 \\
LSTM-l     & 1.03E-03 & 3.80E-04 & 4.91E-04 & 1.23E-01 & 1.16E-02 & 1.31E-03 & 1.36E-05 \\
Ens-Mean-l & 9.57E-04 & 3.42E-04 & 4.62E-04 & 6.60E-02 & 1.13E-02 & 1.19E-03 & 1.31E-05 \\
Ens-Med-l & 9.52E-04 & 3.59E-04 & 5.03E-04 & 5.43E-02 & 1.15E-02 & 1.17E-03 & 1.28E-05 \\
\hline
\multicolumn{8}{|l|}{\textbf{QRS}}\\
\hline
HAR        & 9.85E-04 & \textbf{2.74E-04} & 4.63E-04 & 2.65E-02 & 1.29E-02 & \textbf{1.16E-03} & 1.25E-05 \\
HAR-R      & 9.86E-04 & \textbf{2.74E-04} & 4.44E-04 & 3.01E-02 & 1.29E-02 & \textbf{1.16E-03} & 1.26E-05 \\
ARFIMA     & 9.90E-04 & 3.00E-04 & 4.86E-04 & 3.32E-02 & 1.31E-02 & \textbf{1.16E-03} & 1.22E-05 \\
GARCH      & 1.18E-03 & 3.50E-04 & 5.46E-04 & 7.11E-02 & 1.45E-02 & 1.41E-03 & 1.53E-05 \\
LASSO      & 1.06E-03 & 2.90E-04 & 4.58E-04 & 7.81E-02 & 1.33E-02 & 1.26E-03 & 1.35E-05 \\
RR         & 9.88E-04 & 2.84E-04 & 5.13E-04 & \textbf{2.62E-02} & 1.31E-02 & \textbf{1.16E-03} & 1.21E-05 \\
SVR-G      & 1.01E-03 & 2.96E-04 & 5.47E-04 & 3.32E-02 & 1.35E-02 & 1.19E-03 & \textbf{1.19E-05} \\
SVR-L      & 1.00E-03 & 2.95E-04 & 5.33E-04 & 3.19E-02 & 1.34E-02 & 1.18E-03 & \textbf{1.19E-05} \\
MLP        & 1.03E-03 & 3.03E-04 & 5.34E-04 & 3.44E-02 & 1.37E-02 & 1.20E-03 & \textbf{1.19E-05} \\
FNM        & 1.01E-03 & 2.95E-04 & 4.50E-04 & 3.25E-02 & 1.32E-02 & 1.18E-03 & 1.28E-05 \\
RF         & 1.04E-03 & 3.03E-04 & 5.60E-04 & 3.23E-02 & 1.37E-02 & 1.23E-03 & 1.29E-05 \\
LSTM       & 1.06E-03 & 3.10E-04 & 4.76E-04 & 1.11E-01 & 1.32E-02 & 1.30E-03 & 1.35E-05 \\
Ens-Mean   & 9.83E-04 & 2.78E-04 & 4.43E-04 & 3.21E-02 & 1.29E-02 & \textbf{1.16E-03} & 1.25E-05 \\
Ens-Med    & 9.86E-04 & 2.77E-04 & 4.73E-04 & 2.80E-02 & 1.30E-02 & \textbf{1.16E-03} & 1.24E-05 \\
\hline
\multicolumn{8}{|l|}{\textbf{QLR, l - logaritmized RV, e - extended inputs}}\\
\hline
QLR-l      & 9.57E-04 & 3.48E-04 & \textbf{4.42E-04} & 3.97E-02 & 1.12E-02 & 1.19E-03 & 1.22E-05 \\
QLR-le     & 9.58E-04 & 3.40E-04 & 4.56E-04 & 4.26E-02 & 1.16E-02 & 1.19E-03 & 1.21E-05 \\
QLR        & 9.63E-04 & 3.44E-04 & 4.67E-04 & 3.54E-02 & 1.26E-02 & 1.19E-03 & 1.23E-05 \\
QLR-e      & 9.81E-04 & 3.38E-04 & 5.03E-04 & 5.07E-02 & 1.22E-02 & 1.21E-03 & 1.24E-05 \\
\hline
\multicolumn{8}{|l|}{\textbf{QRF, l - logaritmized RV, e - extended inputs}}\\
\hline
QRF-l      & 9.76E-04 & 3.24E-04 & 5.85E-04 & 3.72E-02 & \textbf{1.11E-02} & 1.23E-03 & 1.27E-05 \\
QRF-le     & 9.78E-04 & 3.23E-04 & 5.82E-04 & 3.93E-02 & 1.13E-02 & 1.23E-03 & 1.28E-05 \\
QRF        & 9.81E-04 & 3.26E-04 & 5.77E-04 & 3.98E-02 & \textbf{1.11E-02} & 1.23E-03 & 1.28E-05 \\
QRF-e      & 9.82E-04 & 3.28E-04 & 5.83E-04 & 4.21E-02 & 1.12E-02 & 1.23E-03 & 1.28E-05 \\ 
\hline
\end{tabular}
\end{center}
\end{table*}

\begin{figure}[t]
\centering
\includegraphics[width=0.35\textwidth]{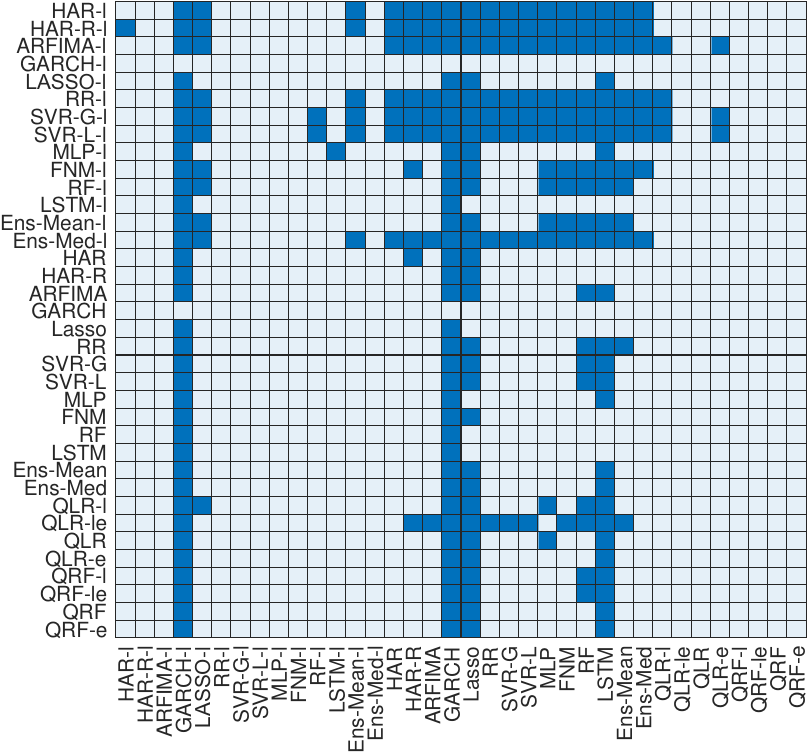}
\caption{Results of the Diebold-Mariano tests for CRPS.} 
\label{figDM}
\end{figure}

To ensure a more robust performance assessment, we conducted a Diebold-Mariano test 
with $\alpha=0.05$ to evaluate the statistical significance of differences in CRPS between each pair of models. The results are visually presented in Fig. \ref{figDM}. A dark square in the diagram indicates that the model on the y-axis is statistically more accurate in terms of CRPS than the model on the x-axis. 
Notably, the best-performing models in this comparison belong to the QRS-l category, including HAR-l, HAR-R-l, AFRIMA-l, RR-l, SVR-G-l, SVR-L-l, and Ens-Med-l. In contrast, GARCH-based models, LASSO, and LSTM perform the worst, as they are outperformed by most other models.

Looking at Table \ref{tab1}, it is worth noting that in the QRS-l variant, individual models exhibit lower mean CRPS than their counterparts in the QRS variant (except for GARCH). However, they have significantly higher median CRPS. This suggests that the QRS variant produces quantile forecasts with greater variance compared to QRS-l, primarily due to the influence of outliers and the asymmetry in the distribution of CRPS values (see the discussion of this phenomenon in the explanation of Fig. \ref{figFor} below). The higher variance in QRS is not evident when examining the IQR of CRPS (see Table \ref{tab1}), as this measure is resistant to outliers (in fact, when comparing corresponding models, the IQR of CRPS is often lower for QRS than for QRS-l).

The QRS ensemble models (Ens-Mean-l, Ens-Med-l, Ens-Mean, and Ens-Med), which aggregate base models, do not significantly improve CRPS compared to the most accurate base models. The same observation holds for more sophisticated ensembling approaches: QLR and QRF.
Among these stacking methods, the variants without extended inputs, operating on logarithmized RV, achieved the lowest mean CRPS. However, QLR demonstrated lower mean CRPS and IQR of CRPS compared to QRF, while QRF exhibited a lower median CRPS. This suggests QRF is more sensitive to outliers than QLR.

Fig. \ref{figFor} presents examples of RV quantile forecasts generated by the QRS, QLR, and QRF approaches. As representatives of QRS, we selected the HAR model in two versions: one using raw RV and the other using logarithmized RV. This model is the simplest among those considered and does not lag behind more complex models in terms of performance metrics. For QLR and QRF, we selected their non-extended versions with logarithmized RV, as they exhibited the lowest mean CRPS values within their respective groups (see Table \ref{tab1}).

\begin{figure}[t]
\centering
\includegraphics[width=0.49\textwidth]{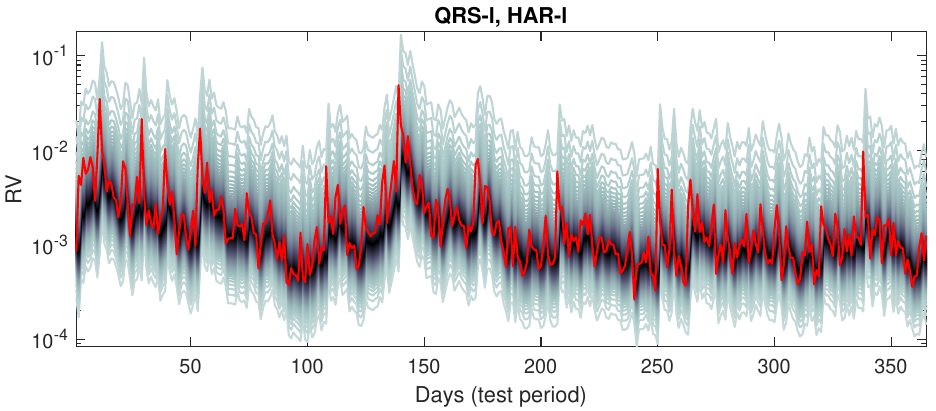}
\includegraphics[width=0.49\textwidth]{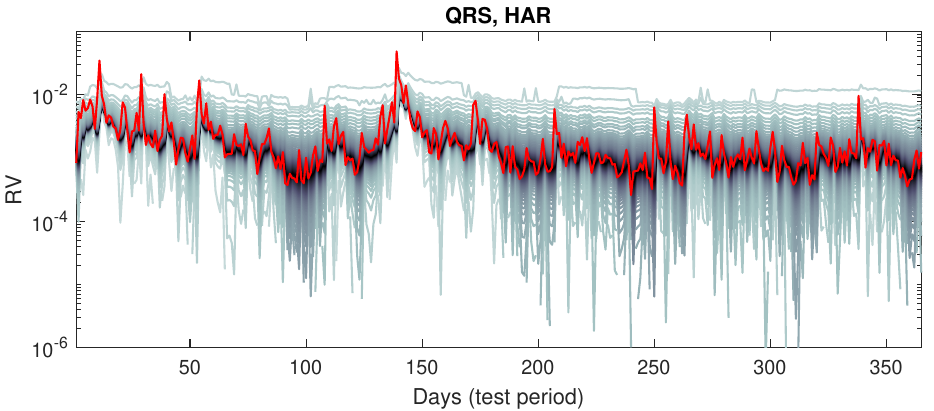}
\includegraphics[width=0.49\textwidth]{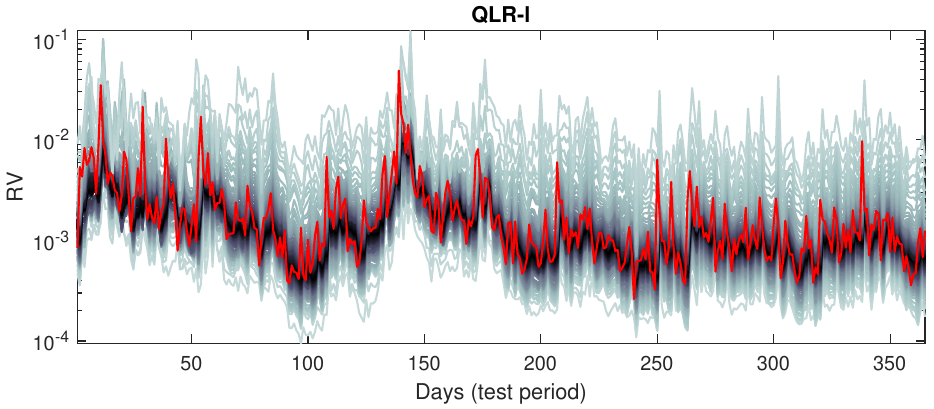}
\includegraphics[width=0.49\textwidth]{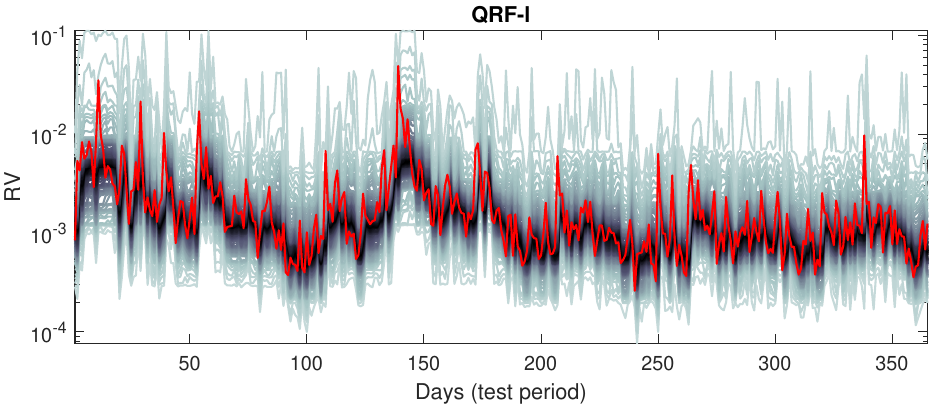}
\caption{Examples of RV quantile forecasts; red line
represents true values.} 
\label{figFor}
\end{figure}

In Fig. \ref{figFor}, quantile forecasts are represented by 99 gray lines, while true values are shown as red lines. A noticeable difference can be observed between the quantiles generated by QRS when applied to raw data versus logarithmized data (top two panels in Fig. \ref{figFor}). In the former case, the lower quantiles are often significantly smaller, sometimes even reaching inadmissible negative values (approximately 3.5\% of cases).
This phenomenon causes the mean CRPS, which is sensitive to outliers, to be higher for QRS than for QRS-l when comparing corresponding models in both variants, despite the median CRPS being lower for QRS.

ML models trained for quantile regression may suffer from quantile crossing, a phenomenon in which a forecast for a lower quantile exceeds that of a higher quantile. This issue is evident in QLR, as shown in Fig. \ref{figFor}, where quantile crossing occurred in nearly 12\% of cases. In contrast, no such occurrences were observed for QRF or QRS, as the methodologies used to construct quantiles in these approaches inherently prevent crossing.

\begin{figure}[t]
\centering
\includegraphics[width=0.24\textwidth]{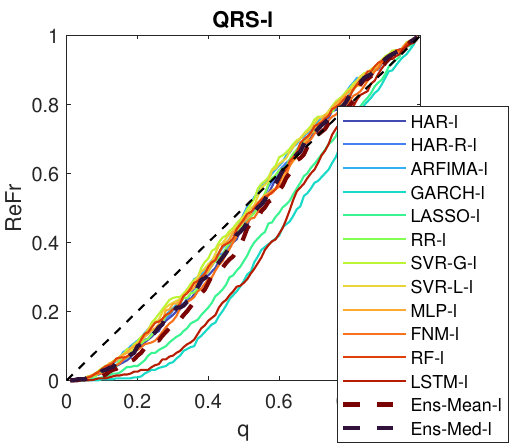}
\includegraphics[width=0.24\textwidth]{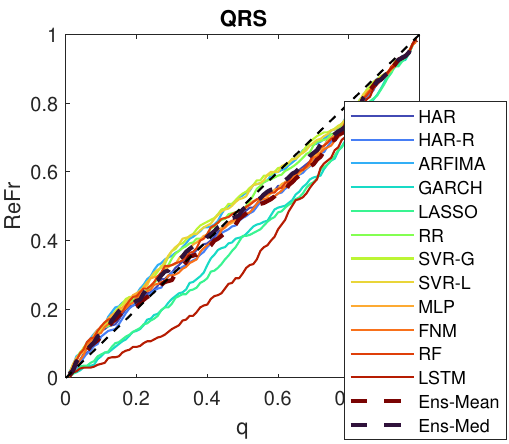}
\includegraphics[width=0.20\textwidth]{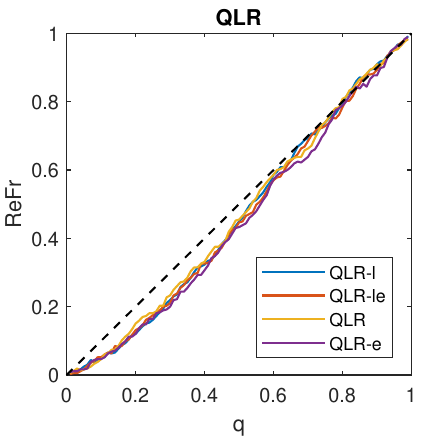}
\includegraphics[width=0.20\textwidth]{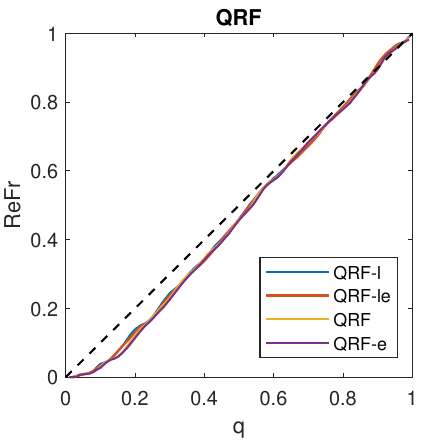}
\caption{Calibration plots.} 
\label{figRF}
\end{figure}

Fig. \ref{figRF} presents the ReFr charts (also called calibration plots or reliability diagrams), with the desired ReFr values indicated by dashed lines. From this figure, it is evident that QRS-l models tend to underestimate lower quantiles (as seen in the downward deviation of the curves from the diagonal for $q < 0.5$) while providing reasonably accurate upper quantiles. A similar pattern is observed for QLR and QRF.

In contrast, for QRS operating on raw RV data (upper right panel), the upper quantiles are underestimated, whereas the lower quantiles are generally well-estimated, except for LSTM, GARCH, and LASSO. The ReFr charts for these three models deviate the most from the diagonal in both QRS variants. This aligns with their significantly higher MARFE values compared to other models, as shown in Table \ref{tab1}.

Notably, QLR variants exhibit similar ReFr charts, while QRF variants are even more closely aligned. Among all models, the QRS approach based on the RR model demonstrates the smallest deviation from the desired ReFr values, with a MARFE of 0.0262 (see Table \ref{tab1}). The next best-performing approach is QRS based on HAR, with a MARFE of 0.0265.

\begin{figure}[t]
\centering
\includegraphics[width=0.22\textwidth]{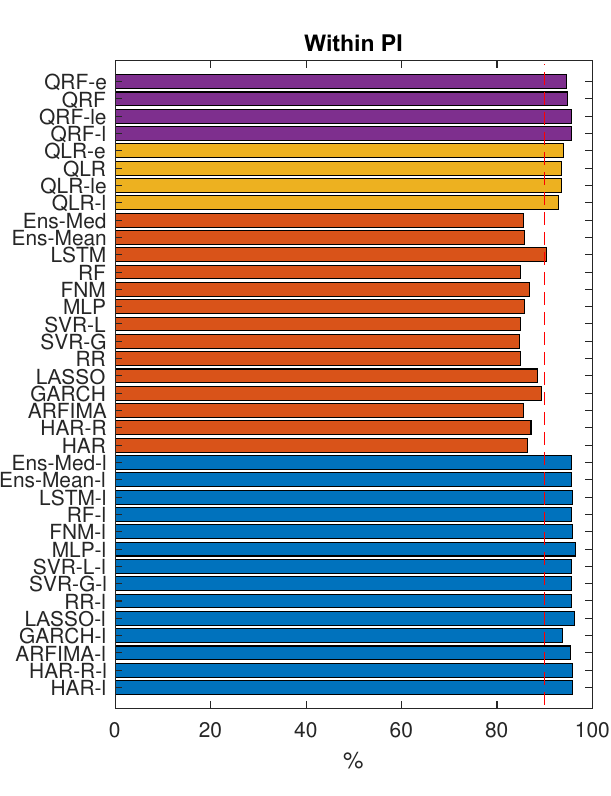}
\includegraphics[width=0.22\textwidth]{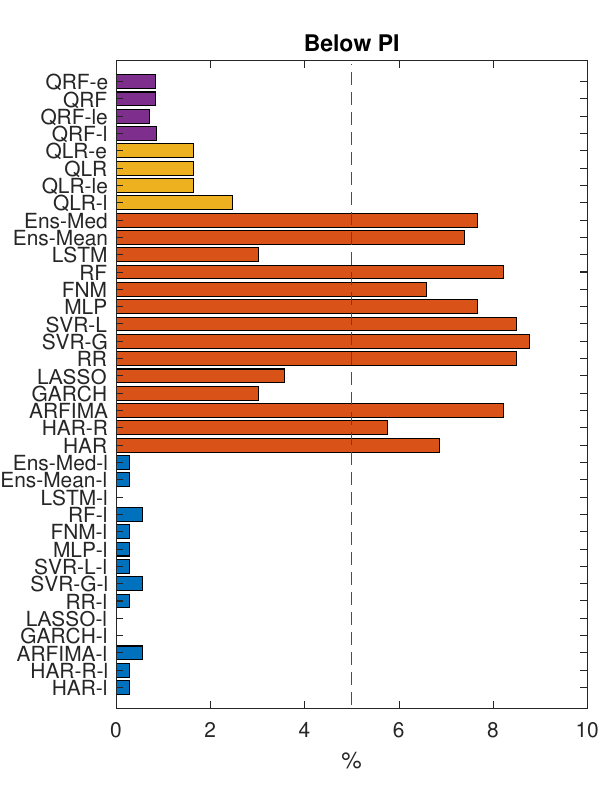}
\includegraphics[width=0.22\textwidth]{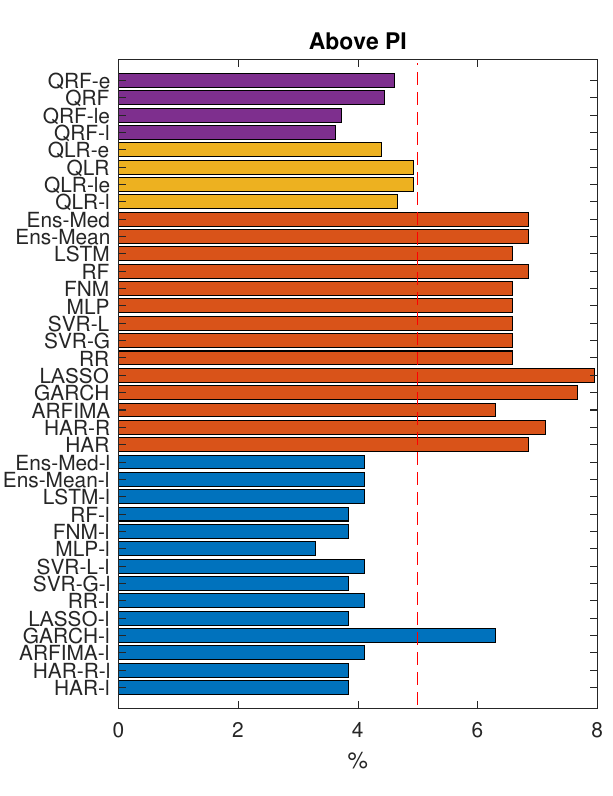}
\caption{Percentage of forecasts within, below and above the 90\% PI. The red dashed line represents the target value.} 
\label{figPI}
\end{figure}

Fig. \ref{figPI} allows us to evaluate the 90\% PIs produced by the models. Notably, the QRS models operating on raw data tend to generate PIs that are mostly too narrow. Among these, the closest to the desired 90\% PI are those from LSTM, GARCH, and LASSO. However, the "Below PI" and "Above PI" plots reveal that these PI intervals are shifted downward, as the number of observations below them is significantly less than the target 5\%, while the number above them is considerably greater than the target 5\%.

The PIs generated by the QRS-l, QLR, and QRF approaches are rated more favorably by the MWS metric compared to QRS (see Table \ref{tab1}). However, these intervals are overly wide and also exhibit a downward shift. 
The highest MWS are observed for the PIs produced by LASSO-l, QRF-l, and QRF. Nevertheless, it is worth noting that many models from the QRS-l and QRF categories achieve MWS close to the best-performing approaches (see Table \ref{tab1}).

Comparing MAE-Q and MSE-Q (Table \ref{tab1}) with the MAE and MSE values obtained from the point forecasts of base models (see \cite{Dud24a}) reveals that, in most cases, the probabilistic approach does not reduce these errors. A slight improvement in MAE was observed only for AFRIMA-l (MAE = 1.17E-3), SVR-G-l (MAE = 1.19E-3), SVR-L-l (MAE = 1.18E-3), and GARCH (MAE = 1.47E-3). Regarding MSE, a reduction was achieved solely for HAR-R (MSE = 1.27E-5). The QLR and QRF approaches exhibit higher MAE-Q and MSE-Q values than the lowest errors obtained by the base models.  
Thus, the probabilistic approach does not improve point forecasts.

QRF utilizes RF, which have built-in mechanisms for estimating the importance of predictors (in this case, the forecasts of the base models). The first method assesses importance using out-of-bag observations (i.e. observations not included in the bootstrap sample) and permuted predictors. For each predictor, importance is determined by measuring the increase in prediction error (MSE) when the predictor's values are randomly permuted across the out-of-bag observations. This calculation is performed for each trained tree, then averaged across the ensemble, and normalized by the standard deviation of the entire ensemble.

The second method estimates importance based on the improvement in the split criterion (MSE) at each split in each tree. The importance measure is then averaged across all trees in the forest for each predictor. As a result, predictors that contribute more significantly to reducing data variability after splits receive higher importance scores.

\begin{figure}[t]
\centering
\includegraphics[width=0.33\textwidth]{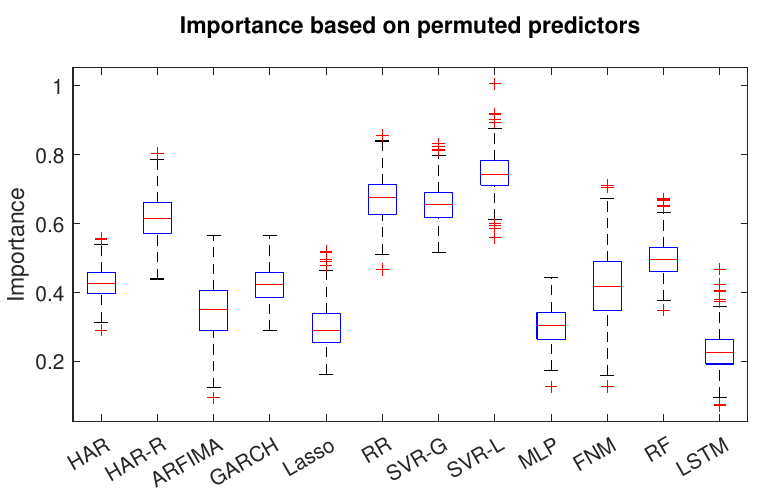}
\includegraphics[width=0.33\textwidth]{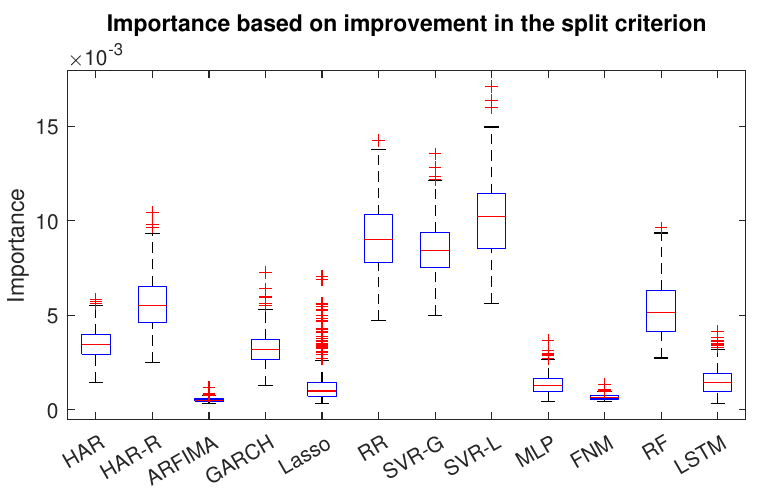}
\caption{Importance of the base models estimated in QRF-l approach.} 
\label{figIM}
\end{figure}

Fig. \ref{figIM} presents the results of both methods for QRF-l. It shows that three base models consistently exhibited the highest importance in both approaches, in the following order: SVR-L, RR, and SVR-G. Regarding the least important models, the method based on permuted predictors identified LSTM, LASSO, and MLP as the least significant, while the method based on the improvement in the split criterion ranked ARFIMA, FNM, LASSO, and MLP as the least influential.

Comparing the quantile forecasting runtimes (for 99 quantiles) on the machine specified above, we obtained approximate execution times of: 0.015 s for QRS, 0.8 s for QLR, and 0.4 s for QRF. These times refer to training and prediction with the meta-models using a dataset of 731 samples with 12 features, and do not include the training time of the base models. Among the methods, QRS is the fastest, as it does not require model training. Its most time-consuming steps are nonparametric kernel density estimation and numerical inversion of CDF.

QRL requires fitting a separate linear model for each quantile level, resulting in computational complexity that scales linearly with the number of quantiles. For example, estimating 99 quantiles entails solving 99 independent linear programming problems. In contrast, QRF estimate the entire quantile distribution in a single training run. The computational cost of QRF primarily depends on the number of trees, their depth, and the size of the dataset. Furthermore, QRF benefits from efficient software implementations and parallel processing, making it more scalable for large datasets.

Although these runtimes are acceptable for daily forecasting applications, they may become a limiting factor in real-time intraday scenarios, where low-latency predictions are essential. In such settings, faster or approximate solutions -- such as reducing the number of quantiles, employing model pruning, or leveraging GPU acceleration -- may be required. Under these conditions, it is difficult to find an alternative as competitive as QRS in terms of speed and efficiency.

\subsection{Discussion}

Our simulations demonstrate that probabilistic stacking can be effectively applied to transform point forecasts into probabilistic ones. Each of the three evaluated approaches has distinct strengths and limitations.

QRS stands out for its conceptual simplicity, computational efficiency, and immunity to quantile crossing, making it particularly well-suited for applications with limited computational resources. However, its reliance on the assumption that past residuals accurately reflect future forecast uncertainty can be problematic, especially under highly volatile market conditions.

QLR, on the other hand, provides a direct and interpretable way to estimate quantiles. Nevertheless, it suffers from quantile crossing in approximately 12\% of forecasts. Moreover, since its loss function is non-quadratic (pinball loss), it requires more computationally expensive optimization (e.g., interior point methods) compared to OLS. Additionally, QLR is inherently limited to modeling linear relationships, which may reduce its effectiveness in capturing complex dependencies.

QRF is well-equipped to model nonlinear relationships and, by design, inherently avoids quantile crossing. Unlike QLR, QRF can simultaneously estimate an arbitrary number of quantiles, making it more efficient for high-resolution probabilistic forecasting. However, the accuracy of quantile estimation is highly dependent on the availability of sufficient training data, as QRF relies on the empirical distribution of observations within the leaf nodes. QRF also has notable limitations: it is computationally intensive and sensitive to hyperparameter settings, particularly the number of trees, splitting criteria, and minimum leaf size, all of which play a crucial role in balancing bias and variance.

Contrary to expectations, incorporating extended inputs, namely daily, weekly, and monthly realized volatility, did not improve predictive performance in the QLR and QRF models. This outcome suggests that the added features may be redundant, as much of this information is likely already embedded in the forecasts generated by the base models. Rather than enhancing the signal, these inputs may introduce noise and increase the risk of overfitting. Nonetheless, this conclusion should not be generalized. In other forecasting tasks or domains, extended inputs could offer complementary information and yield performance gains.

Another noteworthy aspect concerns the use of logarithmic transformation. The results in this regard are mixed. On one hand, comparisons of mean CRPS and MWS values suggest that applying a log transformation often improves forecasting performance by stabilizing variance and reducing skewness in the data. On the other hand, models operating on raw (non-log-transformed) data frequently achieve lower median CRPS and MARFE values. These observations indicate that log transformation tends to reduce sensitivity to outliers, particularly in metrics influenced by extreme values.

Finally, our findings highlight the consistent robustness of the QRS method, particularly when applied to linear base models trained on log-transformed realized volatility. This superior performance likely stems from three factors: (1) the log transformation stabilizes variance and makes the data more amenable to linear modeling; (2) QRS leverages empirical residual distributions, which offer a nonparametric and data-driven representation of uncertainty; and (3) the method’s simplicity reduces the risk of overfitting, particularly in small or noisy datasets. These attributes not only make QRS computationally efficient but also enhance its generalizability to other cryptocurrencies and volatile market conditions. Collectively, our analysis suggests that combining well-specified linear models with residual-based quantile estimation offers a robust, interpretable, and scalable framework for probabilistic volatility forecasting.

\section{Conclusion}

This study addressed the critical gap in cryptocurrency volatility forecasting by proposing and systematically evaluating probabilistic forecasting approaches based on quantile estimation. Recognizing the limitations of traditional deterministic methods, we developed probabilistic forecasts by converting point predictions from multiple base models into quantile-based distributions.  

Our analysis demonstrated that the Quantile Estimation through Residual Simulation method, particularly applied to linear base models using log-transformed realized volatility, consistently yielded robust probabilistic forecasts for Bitcoin. Although sophisticated stacking approaches such as Quantile Linear Regression and Quantile Regression Forests provided valuable quantile forecasts, their overall performance was comparable rather than superior to the simpler QRS-based methods. The probabilistic stacking framework introduced in this research for cryptocurrency volatility forecasting, leveraging QLR and QRF as meta-models, effectively captured the uncertainty and offered reliable volatility estimates, improving upon traditional deterministic stacking approaches. The methodology developed in this study can also be applied to other cryptocurrencies, as their volatility time series share similar characteristics \cite{Dud24a}.

Future research could explore adaptive probabilistic forecasting frameworks that dynamically adjust to varying market volatility regimes, thereby further enhancing predictive stability during extreme market events. Moreover, integrating additional external predictive factors, such as market sentiment or macroeconomic indicators, into the probabilistic stacking framework may improve forecast accuracy. Finally, expanding the proposed methods to multi-step-ahead volatility forecasts or employing deep learning-based probabilistic methods may reveal further performance enhancements and practical advantages in cryptocurrency risk management.

\section*{Acknowledgment}

This study was funded by National Science Centre, Poland (P.F.: grant number 2021/43/B/HS4/00353, W.O.: grant number 2019/35/B/HS4/00642) and Prague University of Economics and Business (P.F.: grant number IP100040).

\vspace{12pt}

\end{document}